# Universal growth scaling law determined by dimensionality

Jinkui Zhao
Songshan Lake Materials Laboratory, Dongguan, Guangdong 523808, China
Institute of Physics, Chinese Academy of Sciences. Beijing 100190, China
Email: jkzhao@iphy.ac.cn

**Growth patterns of complex systems predict how they change in sizes, numbers, masses, etc. Understanding growth is important, especially for many biological, ecological, urban, and socioeconomic systems. One noteworthy growth behavior is the ¾- or and ⅔-power scaling law. It's observed in worldwide aquatic and land biomass productions, eukaryote growth, mammalian brain sizes, and city public facility distributions. Here, I show that these complex systems belong to a new universality class whose system dimensionality determines its growth scaling. The model uses producer-consumer dynamics to derive the $n/(n+1)$ power scaling law for an $n$-dimensional system. Its predictions are validated with real-world two- and three-dimensional data. Dimensionality analysis thus provides a new paradigm for understanding growth and growth-related problems in a wide range of complex systems.**

Growths in complex systems as an emergent property have long been of theoretical interest (*1–5*). Growth behavior is often simple to describe in appearance yet difficult to explain in mechanism. Understanding the growth patterns is especially important to complex biological, ecological, social, and economic systems, particularly because the internal microenvironments and dynamics of these systems are also complex and challenging to explain. Growths in these systems typically follow an empirical power law. In a meta-study of over two thousand ecosystems worldwide in places including grasslands, forests, lakes, and oceans, it was found rather surprisingly that the production of aquatic and land predator biomasses scales close to the ¾- or ⅔-power of prey biomass(*6, 7*). Similar scaling patterns are observed in many other systems: eukaryotes' growth rate(*8*) and mammalian brain size(*9*) scale with the ¾-power of body mass; city public facility density often scales with the ⅔-power of population density(*10, 11*). Each of these empirical laws is significant to its respective field. Understanding biomass production, for example, is important to climate change and renewable energy.

The frequent occurrence of the ¾- and ⅔-power-law growth patterns in such a diverse range of natural and social systems invite questions of whether they belong to a universality class and what their governing mechanism might be. Studies of complex systems benefit from analyzing individual components and their interactions such as with network-based analysis(*12, 13*). However, we are reminded here of P.W. Anderson's insight that 'More is different'(*14*) in that knowledge of system constituents may not be sufficient to explain the emergent properties of a complex system. Anderson's foresight is very relevant to the present biological, ecological, social, and economic systems, which have little in common in their constituents or structures. What these systems do share is their co-existence in the Euclidian space. In this work, I show that it's precisely the system geometry and dimensionality that define their power-law scaling growth patterns.

**Dimensionality determines the growth scaling law**. Growth in a complex system is also the process of resource consumption and conversion. The parts of the system that provide and consume the resources shall be called producers and consumers, respectively. System growth results from the resource flow from the producers to the consumers. During this process, we regard the resources as conserved, namely the resource outflow rate from the producers equals the inflow rate into the consumers. Non-conserved resource flows will not change the discussion as long as resource loss is proportional, for example, if the consumers only convert a fixed percentage of resources into growth. Further, we consider the individual producers and consumers to be invariant during the growth process. Their geometries and sizes remain unchanged, as do their resource production and consumption speeds and rates. An illustrative example of the producer-consumer dynamics is the predator-prey trophic community, where the preys are the producers, and the predators are the consumers. The nominal size of an individual predator and how fast it consumes prey do not change with the system's growth. The question of growth is thus, given a system full of producers and resources, what total size will the consumers grow to.

We now consider an $n$-dimensional system with a characteristic linear dimension $L$ filled with producers. We use mass $M$ to represent the resource. It is proportional to system volume and thus scales with $L$ as $M \sim L^n$. When resources flow out of the producers with the linear speed $u$, the time it takes for the resources to traverse the producers is then $t = L/u$. The mass outflow rate $\Theta$ can thus be written as $\Theta = M/t = Mu/L$.

For the consumers, we view them as a collection of individuals whose numbers will grow. Assuming each consumer has a mass $m_c$ and occupies a space with the characteristic linear dimension $l_c$, the number of consumers that can grow within the system is then $(L/l_c)^n$ and the total consumer mass is $G = (L/l_c)^n m_c$. Like the producers, the linear speed $u_c$ and mass rate $\theta_c$ of resource inflow into an individual consumer are related by $\theta_c = m_c u_c/l_c$. The total mass inflow rate for all the consumers is thus $\Theta_c = (L/l_c)^n m_c u_c/l_c$.

Since the resource flow from the producers to the consumers is conserved, we have $\Theta = \Theta_c$ and $M = G^{(n+1)/n} u_c m_c^{-1/n} u^{-1}$, where we have substituted in the expressions of $\Theta$, $\Theta_c$, and $G$ from the above. Now, because individual consumers and producers are invariant, the individual mass $m_c$, resource inflow speed $u_c$, and resource outflow speed $u$ are all constant. Therefore, $M \sim G^{(n+1)/n}$ and

$$G \sim M^{n/(n+1)}. \tag{1}$$

Equation (1) defines the growth behavior of a new universality class. Growth $G$ scales with the $n/n+1$ power of mass $M$. For the universality class, $M$ is the abstract resource that can be either mass or other proportional quantities in real-world systems. The scaling is sublinear because, with the invariant resource flow speed, the average resource transport time from the producers to the consumers is larger for a larger system, and thus proportionally more resources will spend time in transit. The exponent $n/n+1$ increases with system dimension $n$ because resources flow in more directions from the producers to the consumers in higher-dimensional systems, which means the flow is now more effective.

**Universal growth in two- and three-dimensional systems.** The universal growth scaling law of equation (1) can now be matched and verified with observations. Real-world systems are typically either two- or three-dimensional since we live in a three-dimensional space. For $n = 2$ and 3, the exponent for equation (1) is ⅔ and ¾, respectively. Table 1 lists some of the reported scaling exponents. Plant community biomass productions of grasslands and forests scale with a power exponent of 0.67 against foliage(*6*), which agrees with our prediction for two-dimensional systems. Grasslands and forests are two-dimensional because their heights are negligible compared to their areas. Likewise, African savanna predator versus prey biomass scaling has a power exponent of 0.66(*6, 7*) because animals move in two dimensions.

Aquatic biomes are generally three-dimensional. However, lakes and rivers are often shallower than they are wide and must be viewed as quasi-three-dimensional. If for example, the water depth scales with the square root of lake width $L$, then the system volume and thus system mass will scale with $M \sim L^{2.5}$, and the exponent for equation (1) will be 2.5/3.5 (=0.71). The combined analysis(*6*) of the zooplankton-algal community biomass growth gives a scaling exponent of ~ 0.71, agreeing with this analysis. Dimensionality has been recognized to play an important role in consumer search space in ecosystems(*15, 16*). The current theory offers a first explanation for the wide range of scaling observations in biomass productions(*6*).

Biological systems are typically three-dimensional. It has been known that mammalian brain size scales with the ¾-power of body mass(*9, 17, 18*). The hypothesis is that energy is the main limiting factor as it limits the neuron and brain network formations(*19, 20*). In the current context, the brain is the consumer and the body is the producer that provides the resource in the form of energy. Eukaryote growth is another example, which is reported to scale with the ¾ power of body mass over 20 orders of magnitude of body mass(*6, 8*).

Urban systems are two-dimensional. Minimum facility cost considerations(*21*) and simulations(*12*) also show that they should follow the ⅔-power rule. However, to belong to the universality class defined by equation (1), urban facility growth and distribution must be driven by the population's natural needs such that the preconditions for equation (1) are upheld. The closest such systems are found in public schools(*10*) and post offices(*11*) versus population density scalings. In the current context, the population is viewed as the producers. Their needs and demands are the resources that drive the growth of those public facilities.

**Discussions**

**Out-of-class superlinear scalings.** The invariance preconditions for equation (1) can be satisfied by most natural systems where there are no artificial interventions in their growth dynamics. This may not be the case for many urban and socio-economical. For example, in just-in-time manufacturing, resources can flow more efficiently in bigger cities. Individual productivity also increases when more resources are available. These increased efficiencies violate the consumer-invariance and constant flow speed assumptions. The result can be linear and superlinear scalings, which are often observed in urban, social, and economical settings(*3, 22, 23*). These systems are outside the scope of the new universality class. Their linear and superlinear scalings can be qualitatively illustrated with the following arguments.

For a linear scaling rule, we look into the case where resource flows more efficiently in larger systems such that their flow time from the producers to the consumers remains constant as the system becomes larger. In another word, the time it takes for resources to traverse the producers $t$ and the consumers $t_c$ both remain unchanged. With this new invariance and using the same derivation procedure as for equation (1), we obtain a linear scaling of $G \sim M$. Now, in addition to the linear scaling conditions, we assume resources flow even more efficiently for larger systems and the flow time $t$ diminishes with $t \sim M^{-y}$, $y$ is some arbitrary number, the scaling then becomes superlinear, $G \sim M^{1+y}$. Thus, the higher productivity of denser cities(*3, 22, 23*) can be regarded as a result of more efficient resource transportation and utilization.

**Growth versus metabolism.** The ¾- and ⅔-power growth scaling laws resemble the body-size dependent scalings often observed in biology, the most well-known of which is Kleiber's ¾-power law(*24*) for animal metabolism and its variance of ⅔-power scaling for small

animals(*25*). Metabolism however is the energy expenditure by the whole system, including the producers and consumers in the current context. This distinction provides the needed clarity to ecological discussions such as the "energetic equivalence rule"(*8*). It agrees with the report that across all taxa and over 20 orders of magnitude in body sizes, eukaryotes' metabolism scales isometrically against body mass while their growth obeys the $n/n+1$ scaling law of the current model(*6, 8*). Metabolic scaling only follows power-law approximations within some major groups(*8*) where it is governed by animals' vascular systems that carry oxygen and other nutrients(*26, 27*). Further supporting evidence is found in the observations that community metabolism scales near isometrically against community biomass(*28*) but community biomass growth obeys the present $n/n+1$ scaling law(*6*).

**Further growth-related applications.** There are no known empirical growth observations in one-dimensional systems. Biomass production in a long enough stretch of a river may offer a suitable one-dimensional model. Applications to higher than three dimensions could also be possible with multidimensional physical systems. The current analysis can also help us gain insights into growth-related problems. For cell cultures, cells grown in traditional two-dimensional dishes often behave differently in differentiation, migration, growth, and mechanics from those grown in three dimensions(*29*). Understanding the causes of this difference is important to advancing cell biology and tissue engineering. Various models ranging from biochemical to biomechanical have been put forth to understand how cells distinguish two- from three-dimensional environments(*30*). With the current analysis, growth is affected by dimensionality which in turn can affect the grown cells.

Table 1. Power-law growth scaling exponents for various biological, ecological, and urban systems.

| Complex Systems | Power-law exponent |
|---|---|
| Brain size vs body mass(*9*) | 0.75 |
| Eukaryote growth vs body mass(*6*, *8*) | 0.75 |
| Aquatic predator vs prey biomass(*6*) | 0.71 |
| Land predator vs prey incl. megaherbivores(*6*, *7*) | 0.66 |
| Grassland and forest production vs community biomass(*6*) | 0.67 |
| Public school density vs population(*10*) | 0.69 |
| Post office density vs population(*11*) | 0.7 |